\documentclass[aps,prc,preprint,groupedaddress,showpacs]{revtex4-1}

\usepackage{graphicx}
\usepackage{bm}

\begin{document}


\title{Shell structure in neutron-rich Ca and Ni nuclei
under semi-realistic mean fields}


\author{H. Nakada}
\email[E-mail:\,\,]{nakada@faculty.chiba-u.jp}
\affiliation{Department of Physics, Graduate School of Science,
 Chiba University\\
Yayoi-cho 1-33, Inage, Chiba 263-8522, Japan}


\date{\today}

\begin{abstract}
Shell structure in the neutron-rich Ca and Ni nuclei
is investigated by the spherical Hartree-Fock calculations
with the semi-realistic $NN$ interactions.
Specific ingredients of the effective interaction,
particularly the tensor force, often play a key role
in the $Z$ dependence of the neutron shell structure.
Such examples are found in $N=32$ and $N=40$;
$N=32$ becomes magic or submagic in $^{52}$Ca
while its magicity is broken in $^{60}$Ni,
and $N=40$ is submagic (though not magic) in $^{68}$Ni
but not in $^{60}$Ca.
Comments are given on the doubly magic nature of $^{78}$Ni.
We point out that the loose binding can lead to a submagic number $N=58$
in $^{86}$Ni, assisted by the weak pair coupling.
\end{abstract}


\maketitle



\section{Introduction\label{sec:intro}}
The shell structure,
which is typically manifested in the magic numbers,
is one of the fundamental concepts in the nuclear structure physics.
The shell structure of nuclei is of importance also in astrophysics;
\textit{e.g.} it provides the waiting point of the $s$- and $r$-processes.
As abundant experimental data have been obtained in unstable nuclei,
it has been clarified~\cite{ref:SP08} that
the shell structure may depend on $Z$ or $N$ more strongly
than expected from most conventional theories.
As well as the disappearance of the $N=8$ and $20$ magic numbers,
the new magic numbers $N=16$ and $32$ have been
indicated in neutron-rich nuclei~\cite{ref:N16,ref:N32}.
This discovery has stimulated to reexamine and refine theories
with respect to the nuclear shell structure.
The new experimental facilities~\cite{ref:RIB07}
are expected to access heavier unstable nuclei in coming years.
It is desirable to give predictions on the shell structure
from the refined theories,
which could be a good guidance to new experiments
and will eventually be tested by them.

Concerning the $Z$ or $N$ dependence of the shell structure
(which is sometimes called ``shell evolution''),
two mechanisms have been argued.
The absent or low centrifugal barrier in low-$\ell$ orbits
may influence the shell structure
near the neutron drip line~\cite{ref:N16}.
Whereas the $N=8$ magic number is eroded
because of this mechanism,
there has been no clear evidence for new magic numbers
owing directly to the loose binding.
Since the nuclear shell structure is formed under the average field
composed of the nucleon-nucleon ($NN$) interaction,
the effective $NN$ interaction may also affect the shell structure.
In particular, it has been pointed out that
the tensor force plays a significant role
in the $Z$ or $N$ dependence of the shell structure~\cite{ref:Vtn}.
For full understanding of the shell structure in unstable nuclei,
it will be necessary to take both possibilities into account.

The mean-field (MF) theories provide us with a good tool
to study the nuclear shell structure
from the nucleonic degrees of freedom.
While it is yet difficult to describe
structure of medium- to heavy-mass nuclei
with the fully microscopic $NN$ interaction to good accuracy,
the author has recently developed
semi-realistic $NN$ interactions~\cite{ref:Nak03,ref:Nak08b,ref:Nak10},
in which the Michigan 3-range Yukawa (M3Y) interaction~\cite{ref:M3Y}
is modified so as to reproduce basic observed properties
such as the saturation and the $\ell s$ splitting.
The longest-range part of the central channels
is maintained to be the central force in the one-pion exchange potential
$v^{(\mathrm{C})}_\mathrm{OPEP}$.
The tensor channels in the M3Y-Paris interaction~\cite{ref:M3Y-P}
are contained in the parameter-set M3Y-P5$'$ without any change.
Since the significant part of the tensor force comes from the pions,
M3Y-P5$'$ takes well account of
the leading-order effects of the chiral symmetry breaking.
The tensor channels are dropped in the set M3Y-P4$'$,
which is useful to investigate role of the tensor force.

Shell structure of the neutron-rich Ca and Ni nuclei
is an interesting topic.
The $N=32$ new magic number has been indicated by the experiments
in $^{52}$Ca~\cite{ref:N32,ref:Ca52_Ex2}.
A shell model calculation suggests that magic nature is stronger
in $N=34$ than in $N=32$~\cite{ref:GXPF1A} because of the tensor force,
although the data on $^{56}$Ti show no signature
of the $N=34$ magicity~\cite{ref:Ti56_Ex2}.
Whereas $N=40$ behaves like a magic number in $^{68}$Ni~\cite{ref:Ni68},
contradictory predictions have been given
for $^{60}$Ca~\cite{ref:Nak08b,ref:Nak10,ref:TE06}.
The $Z=28$ magicity has been argued in $^{78}$Ni~\cite{ref:SP08,ref:Vtn}.
It could also be interesting
whether a new magic or submagic number exists
beyond $N=50$ in the Ni isotopes.
In this Communication we shall investigate
shell structure of the neutron-rich Ca and Ni nuclei
by applying the self-consistent Hartree-Fock (HF) calculations
with the semi-realistic $NN$ interactions.

\section{Effective Hamiltonian\label{sec:Hamil}}
Our effective $NN$ interactions have the following form,
\begin{eqnarray} v_{ij} &=& v_{ij}^{(\mathrm{C})}
 + v_{ij}^{(\mathrm{LS})} + v_{ij}^{(\mathrm{TN})}
 + v_{ij}^{(\mathrm{DD})}\,;\nonumber\\
v_{ij}^{(\mathrm{C})} &=& \sum_n \big(t_n^{(\mathrm{SE})} P_\mathrm{SE}
+ t_n^{(\mathrm{TE})} P_\mathrm{TE} + t_n^{(\mathrm{SO})} P_\mathrm{SO}
+ t_n^{(\mathrm{TO})} P_\mathrm{TO}\big)
 f_n^{(\mathrm{C})} (r_{ij})\,,\nonumber\\
v_{ij}^{(\mathrm{LS})} &=& \sum_n \big(t_n^{(\mathrm{LSE})} P_\mathrm{TE}
 + t_n^{(\mathrm{LSO})} P_\mathrm{TO}\big)
 f_n^{(\mathrm{LS})} (r_{ij})\,\mathbf{L}_{ij}\cdot
(\mathbf{s}_i+\mathbf{s}_j)\,,\nonumber\\
v_{ij}^{(\mathrm{TN})} &=& \sum_n \big(t_n^{(\mathrm{TNE})} P_\mathrm{TE}
 + t_n^{(\mathrm{TNO})} P_\mathrm{TO}\big)
 f_n^{(\mathrm{TN})} (r_{ij})\, r_{ij}^2 S_{ij}\,,\nonumber\\
v_{ij}^{(\mathrm{DD})} &=& \big(t_\rho^{(\mathrm{SE})} P_\mathrm{SE}\cdot
 [\rho(\mathbf{r}_i)]^{\alpha^{(\mathrm{SE})}}
 + t_\rho^{(\mathrm{TE})} P_\mathrm{TE}\cdot
 [\rho(\mathbf{r}_i)]^{\alpha^{(\mathrm{TE})}}\big)
 \,\delta(\mathbf{r}_{ij})\,,
\label{eq:effint}\end{eqnarray}
where $\mathbf{r}_{ij}= \mathbf{r}_i - \mathbf{r}_j$,
$r_{ij}=|\mathbf{r}_{ij}|$,
$\mathbf{p}_{ij}= (\mathbf{p}_i - \mathbf{p}_j)/2$,
$\mathbf{L}_{ij}= \mathbf{r}_{ij}\times \mathbf{p}_{ij}$,
$S_{ij}= 4\,[3(\mathbf{s}_i\cdot\hat{\mathbf{r}}_{ij})
(\mathbf{s}_j\cdot\hat{\mathbf{r}}_{ij})
- \mathbf{s}_i\cdot\mathbf{s}_j ]$,
$\hat{\mathbf{r}}_{ij}=\mathbf{r}_{ij}/r_{ij}$,
with $i$ and $j$ representing the indices of nucleons,
and $\rho(\mathbf{r})$ is the nucleon density.
$P_\mathrm{SE}$, $P_\mathrm{TE}$, $P_\mathrm{SO}$ and $P_\mathrm{TO}$
denote the projection operators
on the singlet-even, triplet-even, singlet-odd and triplet-odd
two-particle states.
In the M3Y-type semi-realistic
interactions~\cite{ref:Nak03,ref:Nak08b,ref:Nak10},
the Yukawa function $f_n^{(\mathrm{X})}(r)
=e^{-\mu_n^{(\mathrm{X})} r}/\mu_n^{(\mathrm{X})} r$
is employed ($\mathrm{X}=\mathrm{C}$, $\mathrm{LS}$
and $\mathrm{TN}$).
The density-dependent contact force $v^{(\mathrm{DD})}$
is introduced to realize the saturation.
The parameter-sets M3Y-P4$'$ and P5$'$ are presented
in Ref.~\cite{ref:Nak10}.
We note again that M3Y-P4$'$ contains $v^{(\mathrm{C})}_\mathrm{OPEP}$
but with assuming $v^{(\mathrm{TN})}=0$,
while both $v^{(\mathrm{C})}_\mathrm{OPEP}$ and $v^{(\mathrm{TN})}$
of the M3Y-Paris interaction are untouched in M3Y-P5$'$.
For comparison,
we use the D1S parameter-set~\cite{ref:D1S} of the Gogny interaction,
in which $f_n^{(\mathrm{C})}(r)=e^{-(\mu_n^{(\mathrm{C})} r)^2}$,
the contact form for $v^{(\mathrm{LS})}$,
and $v^{(\mathrm{TN})}=0$ are adopted.

It is reasonably assumed that the spherical symmetry holds
in the neutron-rich Ca and Ni nuclei.
Although the quadrupole deformation cannot always be discarded
for precise studies,
we focus on the spherical shell structure in this Communication,
which is crucial to understand structure of these nuclei.
The spherical HF calculations are implemented
by using the Gaussian expansion method~\cite{ref:NS02,ref:Nak06,
ref:Nak08,ref:NMYM09}
and adopting the Hamiltonian $H=H_N+V_C-H_\mathrm{c.m.}$,
where $H_N (= \sum_i \mathbf{p}_i^2/2M + \sum_{i<j} v_{ij})$,
$V_C$ and $H_\mathrm{c.m.}$ denote the effective nuclear Hamiltonian,
the Coulomb interaction and the center-of-mass Hamiltonian, respectively.
The exchange term of $V_C$ is treated exactly.
Both the one- and the two-body terms of $H_\mathrm{c.m.}$
are subtracted before iteration.

It is noted that, although the D1S interaction does not contain
$v^{(\mathrm{C})}_\mathrm{OPEP}$ and $v^{(\mathrm{TN})}$ explicitly,
a part of their contribution is incorporated in the other channels
in an effective manner.
The same holds for M3Y-P4$'$ that lacks $v^{(\mathrm{TN})}$.
It has still been recognized~\cite{ref:Vtn,ref:Nak08b,ref:Vst,ref:Nak04}
that the $Z$ or $N$ dependence of the shell structure
is difficult to be described without explicit inclusion
of $v^{(\mathrm{C})}_\mathrm{OPEP}$ and $v^{(\mathrm{TN})}$.

\section{Results and discussions\label{sec:result}}
For the Ca and Ni nuclei,
the main correlations beyond the spherical HF solution
should be the neutron pairing.
Therefore the pair energy is a good measure for the neutron shell closure.
The difference between the HF and the Hartree-Fock-Bogolyubov (HFB) energies
has been presented in Refs.~\cite{ref:Nak08b,ref:Nak10}.
In Ref.~\cite{ref:TE06} the neutron pairing gaps have been shown
for the Skyrme energy density functionals SLy4 and SkM$^\ast$.
We further investigate
the shell structure of the neutron-rich Ca and Ni nuclei,
particularly the magic or submagic numbers of $N$,
based on the spherical HF results.

\subsection{Single neutron levels\label{subsec:spe}}
\begin{figure}
\includegraphics[scale=0.9]{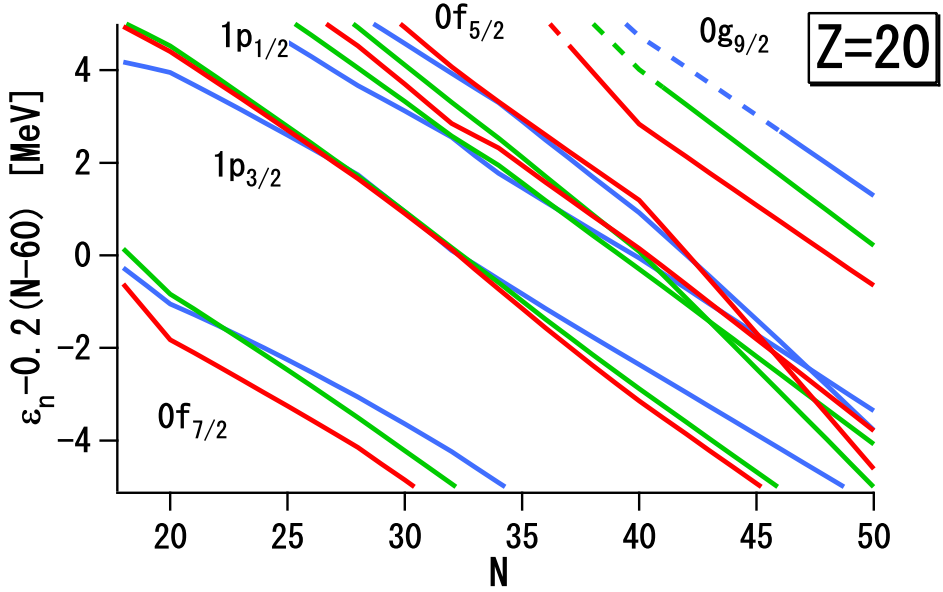}
\caption{$\varepsilon_n(j)$ of the Ca isotopes.
Blue, green and red lines represent the results
with the D1S, M3Y-P4$'$ and P5$'$ interactions, respectively.
Dashed lines are used for positive-energy levels.
\label{fig:Z20_nspe}}
\end{figure}

\begin{figure}
\includegraphics[scale=0.9]{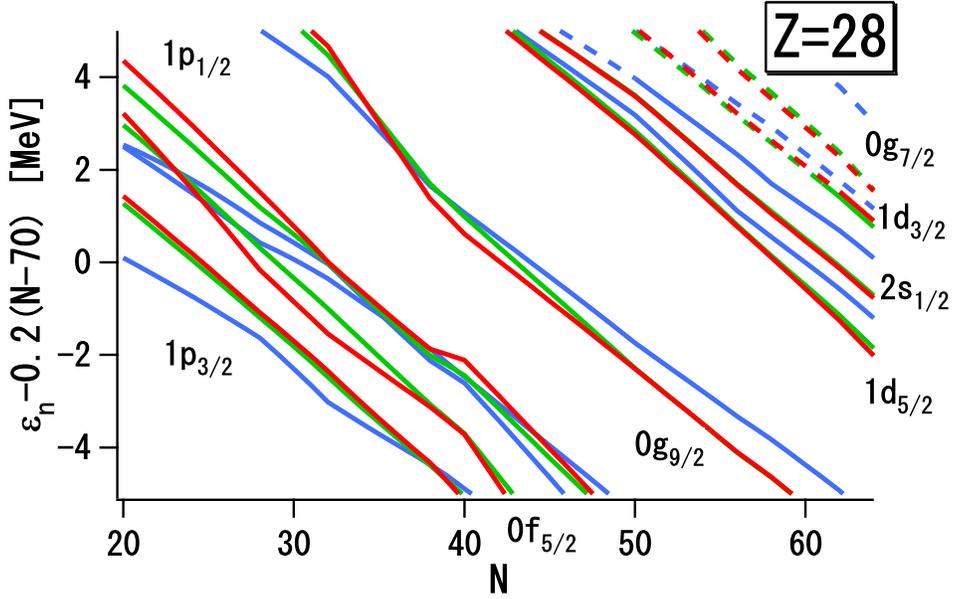}
\caption{$\varepsilon_n(j)$ of the Ni isotopes.
See Fig.~\protect\ref{fig:Z20_nspe} for conventions.
\label{fig:Z28_nspe}}
\end{figure}

The neutron single-particle (s.p.) energies $\varepsilon_n(j)$
around the Fermi level are depicted in Fig.~\ref{fig:Z20_nspe}
for the Ca isotopes,
and in Fig.~\ref{fig:Z28_nspe} for the Ni isotopes.
The HF results obtained from D1S, M3Y-P4$'$ and P5$'$ are compared
with one another.
To keep the figures visible,
$\varepsilon_n(j)$ is shifted by a linear function of $N$
in the vertical axes,
so that the Fermi energies do not largely deviate from the origin.
The dashed lines indicate positive $\varepsilon_n(j)$,
which may correspond to the single neutron resonance
and is shown for reference,
although the correct boundary condition for the resonances
is not taken in the present calculations.

Because of the small difference of the symmetry energy~\cite{ref:Nak10},
the slope of $\varepsilon_n(j)$ in the D1S result is slightly less steep
than those in the M3Y-P4$'$ and P5$'$ results.
A notable point is that $\varepsilon_n(0g_{9/2})$
significantly depends on the interactions in the Ca isotopes,
but not in the Ni isotopes.
It is also noteworthy that the neutron shell structure above $N=50$
in the highly neutron-rich Ni isotopes
is different from the $\beta$ stable region.
The level sequence is $1d_{5/2}$, $2s_{1/2}$, $1d_{3/2}$ and $0g_{7/2}$
from the lower to the higher.

\subsection{$N=32$ and $34$\label{subsec:N32&34}}
\begin{figure}
\includegraphics[scale=0.8]{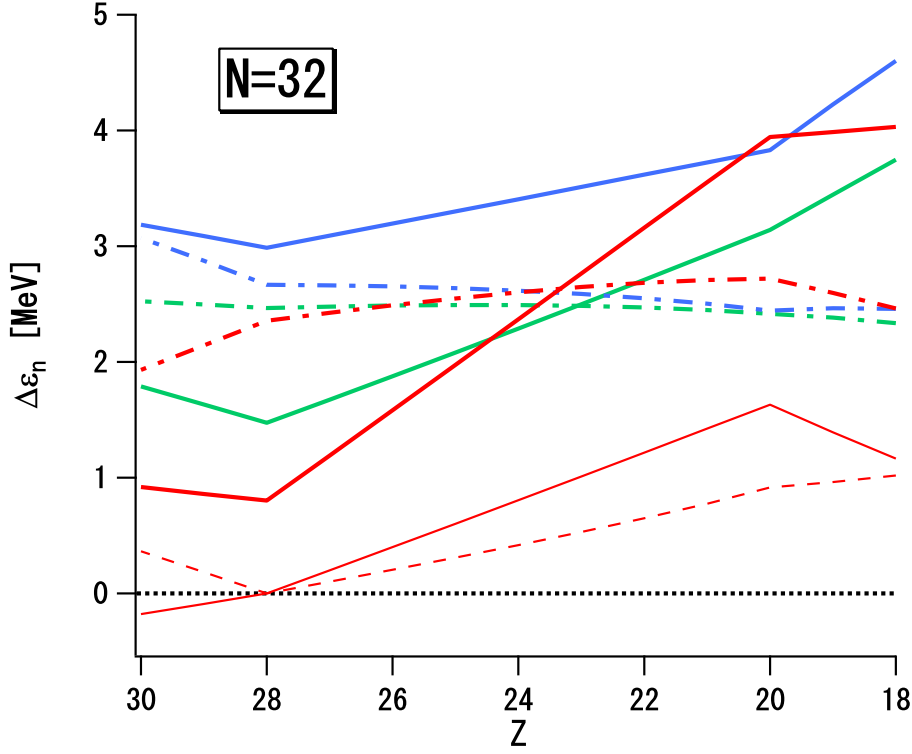}
\caption{$\mathit{\Delta}\varepsilon_n(0f_{5/2})=\varepsilon_n(0f_{5/2})
-\varepsilon_n(1p_{3/2})$ (solid lines)
and $\mathit{\Delta}\varepsilon_n(1p_{1/2})=\varepsilon_n(1p_{1/2})
-\varepsilon_n(1p_{3/2})$ (dot-dashed lines) for the $N=32$ isotones.
See Fig.~\protect\ref{fig:Z20_nspe} for conventions of colors.
Thin red solid and dashed lines represent relative contributions
of $v^{(\mathrm{TN})}$ and $v^{(\mathrm{C})}_\mathrm{OPEP}$
to ${\mathit\Delta}\varepsilon_n(0f_{5/2})$ in the M3Y-P5$'$ result,
after shifting by their values at $^{60}$Ni.
\label{fig:N32_dspe}}
\end{figure}

It is helpful to view $Z$ dependence of the neutron shell structure
in order to pin down what gives rise to the difference
between Ca and Ni.
In Fig.~13 of Ref.~\cite{ref:Nak08b},
the single neutron energies relative to $1p_{3/2}$,
$\mathit{\Delta}\varepsilon_n(j)=\varepsilon_n(j)-\varepsilon_n(1p_{3/2})$,
have been depicted for the $N=32$ isotones as a function of $Z$,
calculated with the interactions M3Y-P4 and P5.
These $\mathit{\Delta}\varepsilon_n(j)$ values are relevant to
the $N=32$ magicity.
We here display $\mathit{\Delta}\varepsilon_n(j)
=\varepsilon_n(j)-\varepsilon_n(1p_{3/2})$ for $j=0f_{5/2}$ and $1p_{1/2}$
calculated with M3Y-P4$'$ and P5$'$ in Fig.~\ref{fig:N32_dspe},
in comparison with those with D1S.
Because of the level inversion,
the $N=32$ shell gap corresponds to $\mathit{\Delta}\varepsilon_n(0f_{5/2})$
in $^{60}$Ni, but to $\mathit{\Delta}\varepsilon_n(1p_{1/2})$ in $^{52}$Ca,
for the M3Y-P4$'$ and P5$'$ results.
In the D1S case $\mathit{\Delta}\varepsilon_n(1p_{1/2})$ represents
the shell gap both in $^{60}$Ni and $^{52}$Ca.
Contributions of $v^{(\mathrm{TN})}$ and $v^{(\mathrm{C})}_\mathrm{OPEP}$
to ${\mathit\Delta}\varepsilon_n(0f_{5/2})$ in the M3Y-P5$'$ result
are also presented.
Because we are interested in the $Z$ dependence
which cannot be compensated by the other channels,
the $v^{(\mathrm{TN})}$ and $v^{(\mathrm{C})}_\mathrm{OPEP}$ contributions
are shifted by their values at $Z=28$.

With the M3Y-P5$'$ interaction we obtain considerable $Z$-dependence
in $\mathit{\Delta}\varepsilon_n(0f_{5/2})$.
This is already recognized by comparing Figs.~\ref{fig:Z20_nspe}
and \ref{fig:Z28_nspe}.
This $Z$-dependence originates in
$v^{(\mathrm{C})}_\mathrm{OPEP}$~\cite{ref:Nak04}
and $v^{(\mathrm{TN})}$~\cite{ref:Nak08b},
both of which act attractively on $n0f_{5/2}$ as $p0f_{7/2}$ is occupied,
via the mechanism discussed in Refs.~\cite{ref:Vtn,ref:Vst}.
Note that, though the same mechanism is present also for $n1p_{1/2}$,
the effects are much smaller and do not lead to significant $Z$ dependence.
Not including these parts explicitly, the D1S interaction does not provide
strong $Z$-dependence in ${\mathit\Delta}\varepsilon_n(0f_{5/2})$.
Having $v^{(\mathrm{C})}_\mathrm{OPEP}$ but not $v^{(\mathrm{TN})}$,
M3Y-P4$'$ gives moderate $Z$-dependence.
The shell gaps in $^{52}$Ca are comparable among all the interactions.
The pair energies shown in Refs.~\cite{ref:Nak10,ref:TE06}
confirm that $^{52}$Ca is nearly a doubly-magic nucleus,
as is consistent with the measured $E_x(2^+_1)$ value~\cite{ref:Ca52_Ex2}.
On the contrary, the experimental data in $^{60}$Ni show no enhancement
of $E_x(2^+_1)$~\cite{ref:TI},
suggesting meltdown of the $N=32$ magicity.
The effects of the tensor force on the mean fields,
together with those of $v^{(\mathrm{C})}_\mathrm{OPEP}$,
well account for the $Z$-dependence of the $N=32$ magicity.

Unlike the $N=32$ shell gap,
the difference between $\varepsilon_n(0f_{5/2})$ and $\varepsilon_n(1p_{1/2})$
is not remarkable at $Z=20$ in the present calculations.
As a result $^{54}$Ca has a certain amount of the pair excitation,
as shown in Ref.~\cite{ref:Nak10}.
It is emphasized that this consequence is obtained
even with M3Y-P5$'$ that includes reasonably strong tensor force.
Thus the $N=34$ magicity cannot be concluded only from the tensor force,
and influence of the other parts of the interaction
(\textit{e.g.} the central channels)
on $\varepsilon_n(0f_{5/2})-\varepsilon_n(1p_{1/2})$ is important as well.

\subsection{$N=40$\label{subsec:N40}}
\begin{figure}
\includegraphics[scale=0.9]{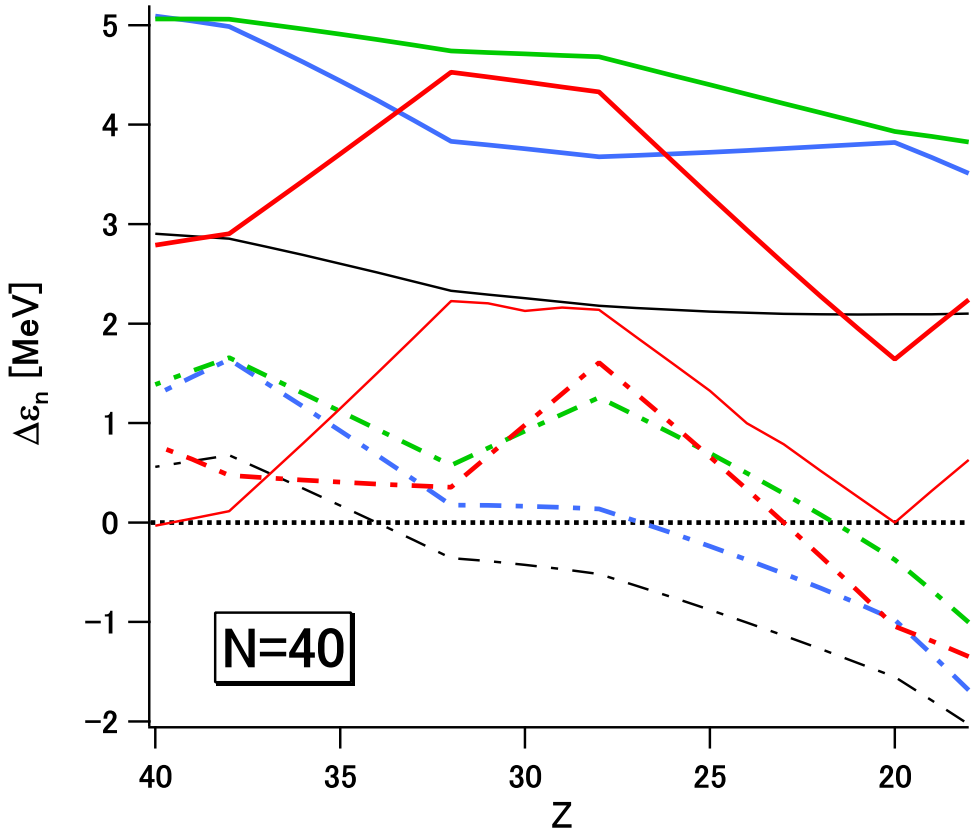}
\caption{$\mathit{\Delta}\varepsilon_n(0g_{9/2})=\varepsilon_n(0g_{9/2})
-\varepsilon_n(0f_{5/2})$ (solid lines)
and $\mathit{\Delta}\varepsilon_n(1p_{1/2})=\varepsilon_n(1p_{1/2})
-\varepsilon_n(0f_{5/2})$ (dot-dashed lines) for the $N=40$ isotones.
For thick lines, the same colors as in Fig.~\protect\ref{fig:Z20_nspe}
are used.
Thin black lines are the results of the SkM$^\ast$ interaction.
Relative contribution of $v^{(\mathrm{TN})}$
to ${\mathit\Delta}\varepsilon_n(0g_{9/2})$ in the M3Y-P5$'$ result
is shown by the thin red line,
with shifting by its value at $^{60}$Ca.
\label{fig:N40_dspe}}
\end{figure}

As viewed in Refs.~\cite{ref:Nak08b,ref:Nak10,ref:TE06},
the magic or submagic nature of $N=40$ predicted by the MF calculations
significantly depends on the input effective interactions.
While the pair excitation is hindered both in $^{68}$Ni and $^{60}$Ca
if we use SLy4, D1S or M3Y-P4$'$,
there is no signature of the $N=40$ magicity with SkM$^\ast$.
All of these interactions do not contain the explicit tensor force.
If we apply M3Y-P5$'$ that contains realistic tensor force,
the pair excitation is highly suppressed in $^{68}$Ni but not in $^{60}$Ca.
These results are traced back to the s.p. energy of $n0g_{9/2}$
relative to $n0f_{5/2}$ and $n1p_{1/2}$.
We present the $Z$ dependence of
$\mathit{\Delta}\varepsilon_n(j)=\varepsilon_n(j)-\varepsilon_n(0f_{5/2})$
($j=0g_{9/2}$ and $1p_{1/2}$) for the $N=40$ isotones
in Fig.~\ref{fig:N40_dspe}.
The $N=40$ shell gap is represented by $\mathit{\Delta}\varepsilon_n(0g_{9/2})$
when $\mathit{\Delta}\varepsilon_n(1p_{1/2})$ is negative,
and by $\mathit{\Delta}\varepsilon_n(0g_{9/2})
-\mathit{\Delta}\varepsilon_n(1p_{1/2})$
when $\mathit{\Delta}\varepsilon_n(1p_{1/2})$ is positive.

As mentioned above,
the $\mathit{\Delta}\varepsilon_n(0g_{9/2})$ values
obtained with D1S, M3Y-P4$'$ and SkM$^\ast$ do not strongly depend on $Z$.
It has experimentally been suggested that $^{68}$Ni looks like
a doubly magic nucleus~\cite{ref:Ni68}.
The D1S interaction, which gives strongly suppressed pairing,
describes $E_x(2^+_1)$ and $B(E2)$ of $^{68}$Ni to good accuracy
within the quasiparticle version
of the random-phase approximation (RPA)~\cite{ref:Peru09}.
The sizable pair excitation with SkM$^\ast$
is ascribed to the small $\mathit{\Delta}\varepsilon_n(0g_{9/2})$.
It would be difficult to reproduce the experimental data
on $E_x(2^+_1)$ and $B(E2)$ in $^{68}$Ni
with this small $\mathit{\Delta}\varepsilon_n(0g_{9/2})$.
If we use M3Y-P5$'$ that includes realistic tensor force,
the $N=40$ shell gap is comparable to those with D1S and M3Y-P4$'$
in $^{68}$Ni,
but $\mathit{\Delta}\varepsilon_n(0g_{9/2})$ significantly depends on $Z$.
This $Z$-dependence is predominantly carried by the tensor force,
as clarified in Fig.~\ref{fig:N40_dspe}.
Contribution of $v^{(\mathrm{C})}_\mathrm{OPEP}$ is not important in this case.
The $N=40$ shell gap comes minimum at $Z=20$,
and this leads to sizable pair excitation in $^{60}$Ca~\cite{ref:Nak10}.
As long as we rely on the shell gap in $^{68}$Ni,
it is likely that the $N=40$ magicity is significantly broken in $^{60}$Ca,
since the tensor force certainly exists in the $NN$ interaction.
This prediction will be tested by future experiments
on systematics of the binding energies
and/or of the first excited states.

There have been arguments on the $N=40$ magicity
in $^{68}$Ni~\cite{ref:N40,ref:SLD02,ref:LTN03}.
Although the pair energy is quite small,
the neutrons are still in the superfluid phase
in the HFB results of $^{68}$Ni, with any of D1S, M3Y-P4$'$ and M3Y-P5$'$.
Moreover, the experimental data show that the magicity is lost quickly
as $Z$ departs from $28$~\cite{ref:TI}.
This situation reminds us of the protons in $^{146}$Gd~\cite{ref:MNOM},
and it could be more reasonable to call $N=40$ around $^{68}$Ni
a \textit{submagic} number rather than a new magic number.

A small shell gap may induce quadrupole deformation.
The quadrupole deformation in the $N=40$ isotones,
as is known for $^{80}$Zr,
has been investigated in Ref.~\cite{ref:N40},
by using the D1S interaction.
Though beyond the scope of this paper,
it will be of interest to study deformation effects
with the semi-realistic interactions.


\subsection{$^{78}$Ni\label{subsec:N50}}
In the present MF calculations,
the magic number $N=50$ is maintained both for $^{70}$Ca and $^{78}$Ni,
irrespective of the effective interactions.
The shell gap between $n0g_{9/2}$ and the upper levels
is large enough to prevent the neutrons from being excited
in the HFB calculations.

It has been suggested~\cite{ref:SP08,ref:Vtn} that
the $Z=28$ magic nature could be eroded in $^{78}$Ni,
because $p0f_{5/2}$ comes down via the attraction
from the protons occupying $p0f_{7/2}$.
In the HF calculation with M3Y-P5$'$,
such attraction is realized because the tensor force is included,
and $p0f_{5/2}$ becomes the lowest unoccupied proton orbit.
However, it is not sufficient to violate the $Z=28$ shell gap,
which amounts to $5.8\,\mathrm{MeV}$, via the pair excitation.
Since the magic nature is usually linked to
properties of the first excited state,
we present the values of $E_x(2^+_1)$
and $B(E2)$ predicted by the HF+RPA calculations in Table~\ref{tab:Ni78-2+}.
As well as those of D1S and the semi-realistic interactions,
the results of the new parameter-set of the Gogny interaction
D1M~\cite{ref:D1M} are displayed.
Comparison with future experiments is desired.

\begin{table}
\begin{center}
\caption{$E_x(2^+_1)$ and $B(E2;2^+_1\rightarrow 0^+_1)$ in $^{78}$Ni,
 predicted by the HF+RPA calculations.
\label{tab:Ni78-2+}}
\begin{tabular}{ccrrrr}
\hline\hline
&&~~~~~D1S~~&~~~~~D1M~~&~~M3Y-P4$'$&~~M3Y-P5$'$\\ \hline
$E_x(2^+_1)$ &(MeV)& $3.15$ & $3.00$ & $3.28$ & $3.25$ \\
$B(E2;2^+_1\rightarrow 0^+_1)$ &($e^2\mathrm{fm}^4$)&
 $84.4$ & $83.4$ & $87.6$ & $84.4$ \\
\hline\hline
\end{tabular}
\end{center}
\end{table}

\subsection{$N=58$\label{subsec:N58}}
\begin{figure}
\includegraphics[scale=0.9]{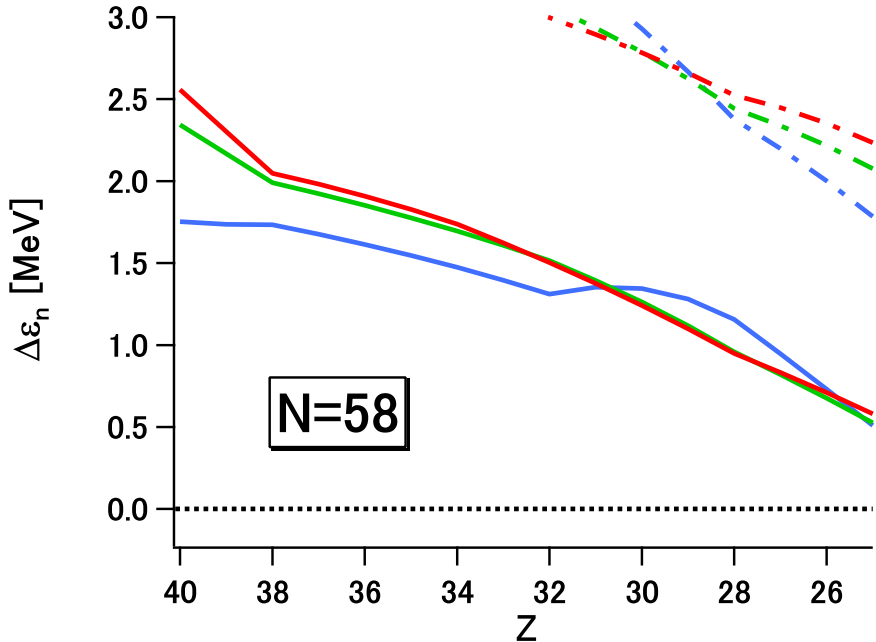}
\caption{$\mathit{\Delta}\varepsilon_n(2s_{1/2})=\varepsilon_n(2s_{1/2})
-\varepsilon_n(1d_{5/2})$ (solid lines)
and $\mathit{\Delta}\varepsilon_n(1d_{3/2})=\varepsilon_n(1d_{3/2})
-\varepsilon_n(1d_{5/2})$ (dot-dashed lines) for the $N=58$ isotones.
See Fig.~\protect\ref{fig:Z20_nspe} for conventions of colors.
\label{fig:N58_dspe}}
\end{figure}

References~\cite{ref:Nak10,ref:TE06} show
that the pair correlation is greatly suppressed in $^{86}$Ni,
suggesting the submagic nature of $N=58$ near the neutron drip line.
To examine the neutron shell structure around $N=58$,
$\mathit{\Delta}\varepsilon_n(j)=\varepsilon_n(j)-\varepsilon_n(1d_{5/2})$
is depicted for $j=2s_{1/2}$ and $1d_{3/2}$ in Fig.~\ref{fig:N58_dspe}.
Basically the interval between $2s_{1/2}$ and $1d_{3/2}$ corresponds to
the $N=58$ shell gap.
However, being mostly positive
while not satisfying the correct boundary condition,
$\varepsilon_n(1d_{3/2})$ would not precisely represent resonances.
Nevertheless the calculated energies of $n1d_{3/2}$ are useful
in interpreting the current HFB and RPA results of $^{86}$Ni,
in which influence of the continuum is
efficiently taken into account~\cite{ref:Nak06,ref:NMYM09}.

In this region we do not find remarkable interaction-dependence
in the neutron shell structure.
As approaching the neutron drip line (\textit{i.e.} for decreasing $Z$),
the lower-$\ell$ orbit has relatively lower energy
because its wave function feels the weaker centrifugal repulsion
and thereby easily extends in the coordinate space.
The main correlation which may break the $N=58$ shell gap
is the pair excitation out of $n2s_{1/2}$ to $n1d_{3/2}$.
The coupling between these two orbits via the pairing is not strong,
primarily because their degeneracy ($2j+1$) is small.
Indeed, the coupling matrix element
$\langle(n1d_{3/2})^2\,J=0|v_{ij}|(n2s_{1/2})^2\,J=0\rangle$
is $\approx 0.3\,\mathrm{MeV}$
if evaluated by the M3Y-P5$'$ interaction,
appreciably smaller than
$2[\varepsilon_n(1d_{3/2})-\varepsilon_n(2s_{1/2})]\approx 3\,\mathrm{MeV}$.
Thus, the loose binding, assisted by the weak coupling,
leads to the $N=58$ submagic nature in $^{86}$Ni,
although the pair excitation remains within the HFB regime.

Because the weak coupling plays a certain role,
the submagic number $N=58$ at $^{86}$Ni does not imply high $E_x(2^+_1)$.
On the other hand, it could be manifested
by suppressed $B(E2;2^+_1\rightarrow 0^+_1)$.
The $2^+_1$ state is not easily handled in numerical calculations,
because it is located just above the neutron threshold.
Though the boundary condition should be treated more carefully
to get precise values,
we note for reference that the HF+RPA calculations
using the Gaussian expansion method~\cite{ref:NMYM09}
give $E_x(2^+_1)\sim 1\,\mathrm{MeV}$ and
$B(E2;2^+_1\rightarrow 0^+_1)=10-20\,e^2\mathrm{fm}^4$.

\section{Summary\label{sec:summary}}
We have investigated shell structure of the neutron-rich Ca and Ni nuclei
by the spherical Hartree-Fock calculations
mainly with the semi-realistic $NN$ interaction M3Y-P5$'$.
In $Z$ dependence of the neutron magic or submagic numbers,
specific ingredients of the effective interaction,
particularly the tensor force, could play a crucial role.
The magic nature of $N=32$ around $^{52}$Ca
and the non-magic nature around $^{60}$Ni can be accounted for
by the tensor force as well as by the central part
of the one-pion exchange potential;
\textit{i.e.} the leading order effects of the chiral symmetry breaking.
On the other hand, the present mean-field study
does not support the $N=34$ magic number.
The tensor force gives rise to $Z$ dependence of the shell structure
around $N=40$.
Whereas the submagic nature of $N=40$ in $^{68}$Ni has been observed
and is described by many effective interactions except SkM$^\ast$,
the submagic nature is likely destroyed in $^{60}$Ca
because of the $Z$ dependence in the shell structure
produced by the tensor force.

Although it has been pointed out that
the loose binding could lead to new magic numbers in drip-line nuclei,
no clear evidence has been found so far.
We point out that $N=58$ will be submagic in $^{86}$Ni,
owing to the lower centrifugal barrier in the lower $\ell$ orbits,
together with the weak pair coupling.
This submagic nature may be connected to small $B(E2)$,
but not to high $E_x(2^+_1)$.

In the present work we have constrained ourselves
to the spherical MF calculations,
which are useful to understand variation of the structure
in the Ca to Ni nuclei in a simple manner.
Future plan includes extension of the calculations
by taking the possibility of deformation into consideration~\cite{ref:Nak08},
as has been done with the phenomenological Skyrme or Gogny
energy density functionals~\cite{ref:N40,ref:TIJ2}.

\begin{acknowledgments}
%
This work is financially supported in part
as Grant-in-Aid for Scientific Research (C), No.~22540266,
by Japan Society for the Promotion of Science.
Numerical calculations are performed on HITAC SR11000
at Institute of Media and Information Technology, Chiba University,
at Information Technology Center, University of Tokyo,
and at Information Initiative Center, Hokkaido University.
The code given in Ref.~\cite{ref:CNP1} is employed
for the calculation with SkM$^\ast$.
\end{acknowledgments}


\end{document}